\documentclass[conference]{IEEEtran}
\IEEEoverridecommandlockouts

\usepackage{amsmath,amssymb,amsfonts}
\usepackage{algorithmic}
\usepackage{algorithm}
\usepackage{array}
\usepackage{textcomp}
\usepackage{stfloats}
\usepackage{url}
\usepackage{verbatim}
\usepackage{graphicx}
\usepackage{cite}
\usepackage{subcaption}
\def\BibTeX{{\rm B\kern-.05em{\sc i\kern-.025em b}\kern-.08em
		T\kern-.1667em\lower.7ex\hbox{E}\kern-.125emX}}

\makeatletter
\newcommand{\linebreakand}{%
\end{@IEEEauthorhalign}
\hfill\mbox{}\par
\mbox{}\hfill\begin{@IEEEauthorhalign}
}
\makeatother

\begin{document}

\title{\huge Supporting Urban Low-Altitude Economy: Channel Gain Map Inference Based on 3D Conditional GAN}

\author{ 
	\IEEEauthorblockN{ Yonghao Wang\IEEEauthorrefmark{1}, Ruoguang Li\IEEEauthorrefmark{1}, Di Wu\IEEEauthorrefmark{2}\IEEEauthorrefmark{3}, Jiaqi Chen\IEEEauthorrefmark{1}, and Yong Zeng\IEEEauthorrefmark{2}\IEEEauthorrefmark{3}}
	\IEEEauthorblockA{\IEEEauthorrefmark{1} College of Information Science and Engineering, Hohai University, Changzhou 213200, China\\}
	\IEEEauthorblockA{\IEEEauthorrefmark{2} National Mobile Communications Research Laboratory, Southeast University, Nanjing 210096, China\\}
	\IEEEauthorblockA{\IEEEauthorrefmark{3} Purple Mountain Laboratories, Nanjing 211111, China\\}
	Email: \{yonghaowang, ruoguangli, jiaqichen\}@hhu.edu.cn, \{studywudi, yong\_zeng\}@seu.edu.cn
\thanks{
This work was supported in part by the National Natural Science Foundation of China under Grant 62301157, in part by the Natural Science Foundation of Jiangsu Province of China under Project BK20230823. }
}
\maketitle

\begin{abstract}
The advancement of advanced air mobility (AAM) in recent years has given rise to the concept of low-altitude economy (LAE). However, the diverse flight activities associated with the emerging LAE applications in urban scenarios confront complex  physical environments, which urgently necessitates ubiquitous and reliable communication to guarantee the operation safety of the low-altitude aircraft. As one of promising 
technologies for the sixth generation (6G) mobile networks, channel knowledge map (CKM) enables the environment-aware communication by constructing a site-specific dataset, thereby providing a priori on-site information for the aircraft to obtain the channel state information (CSI) at arbitrary locations with much reduced online overhead. Diverse base station (BS) deployments in the three-dimensional (3D) urban low-altitude environment require efficient 3D CKM construction to capture spatial channel characteristics with less overhead. Towards this end, this paper proposes a 3D channel gain map (CGM) inference method based on a 3D conditional generative adversarial network (3D-CGAN). Specifically, we first analyze the potential deployment types of BSs in urban low-altitude scenario, and investigate the CGM representation with the corresponding 3D channel gain model. The framework of the proposed 3D-CGAN is then discussed, which is trained by a dataset consisting of existing CGMs. Consequently, the trained 3D-CGAN is capable of inferring the corresponding CGM only based on the BS coordinate without additional measurement. The simulation results demonstrate that the CGMs inferred by the proposed 3D-CGAN outperform those of the benchmark schemes, which can accurately reflect the radio propagation condition in 3D environment.

\end{abstract}

\section{Introduction}
The proliferation of the crewed and uncrewed aircraft in recent years, such as unmanned aerial vehicles (UAVs) and electric vertical take-off and landing (eVTOL), has accelerated the development of advanced air mobility (AAM), which, in turn, brings about a new economic paradigm named low-altitude economy (LAE) \cite{ref1}. LAE is envisioned to exploit airspace below traditional aviation altitudes (typically below 1,000 meters or 3,000 feet) for a variety of industrial and public service applications, such as emergency response, logistics, and urban air transportation \cite{ref1-1}. For all the aforementioned low-altitude operations to function smoothly, ubiquitous and reliable communication plays a key role to enable seamless and real-time aircraft coordination and control\cite{ref1-2}. However, unlike the rural area with relatively open space, the low-altitude airspace in the urban area confronts complex physical environments due to the existence of large number of buildings with different heights. This issue may force the base station (BS) to carry out frequent channel state information (CSI) acquisition to meet the link robustness, leading to a significant increase of system overhead. 

As one of promising technologies for the sixth generation (6G) mobile networks, channel knowledge map (CKM) was recently proposed in \cite{ref2} to enable the environment-aware communication by constructing a site-specific dataset encompassing the location information of transmitters and receivers. Various channel knowledge has been considered, such as channel gains, channel angle information, and beam index\cite{ref3}\cite{ref4}.  In contrast to the radio environment map (REM)\cite{REM}, CKM emphasizes the direct representation of the intrinsic wireless channel characteristics at particular locations. Specifically, channel gain map (CGM) is a specific type of CKM that provides the channel gain knowledge at any locations within a certain region\cite{ref5}. Therefore, by leveraging CKM, the aircraft can obtain a priori on-site CSI at arbitrary unknown locations\cite{streaming}, which supports the development of urban LAE from a wireless perspective.

Existing works for constructing CKM can be roughly categorized into four types: model-based, interpolation, ray-tracing-based, and machine learning methods\cite{ref6}. The authors in \cite{ref7} compared model-based and model-free methodologies for constructing CGM, extending the current model-based channel prediction framework through the incorporation of a multi-mode channel model. In \cite{ref8}, the CKM construction process was formulated as an image-to-image (I2I) inpainting task, which was solved by a Laplacian Pyramid (LP)-based approach. The authors in \cite{ref9} proposed a two-stage interference-aware CKM construction method consisting of the interfering signal strength (ISS) map and the signal-to-interference-plus-noise ratio (SINR) map. In \cite{Unet}, the inferred CKMs of potentially new access points (APs) can be generated across APs with the trained UNet and the channel knowledge of the
existing APs. It can be seen that most of the current literature focuses on the two-dimensional (2D) CKM construction, which cannot properly reflect the true wireless environment in three-dimensional (3D) region in urban low-altitude scenarios. Recently, several works have made efforts on 3D CKM construction. Specifically, \cite{ref10} proposed a Kriging interpolation-based 3D CKM construction scheme with the assistance of UAV measurement. The authors in \cite{ref11} selected a certain number of discrete layers to represent the UAV flight region and employed extreme gradient boost (XGBoost) to obtain the channel condition in advance. However, such a layering approach may cause a channel knowledge blind zone for the aircraft that continuously traverse different altitudes. Additionally, both of the above works have to use instantaneous measurement data during CKM construction. This implies that the measurement samples must be recollected once the BS location is changed, resulting in a significant overhead.

In order to address the above issues, this paper investigates a 3D CGM inference method based on a 3D conditional generative adversarial network (3D-CGAN) for a urban low-altitude scenario. The proposed method is able to infer the CGM for a new location by leveraging existing CGMs without additional measurement sampling. Specifically, we first analyze the potential deployment types of BSs in the urban low-altitude scenario, and then investigate the CGM representation with the corresponding 3D channel gain model. The framework of the 3D CGM inference by the proposed 3D-CGAN is discussed. By utilizing existing CGMs for training, the trained model can infer the corresponding CGM only based on the BS coordinate without requiring any additional samplings. The simulation results demonstrate that the CGMs inferred by the proposed scheme outperform those inferred by the benchmark schemes, which can accurately reflect the radio propagation condition in 3D environment.

\section{System Model}

\begin{figure}[t]
	\centering
	\includegraphics[width=3.2in]{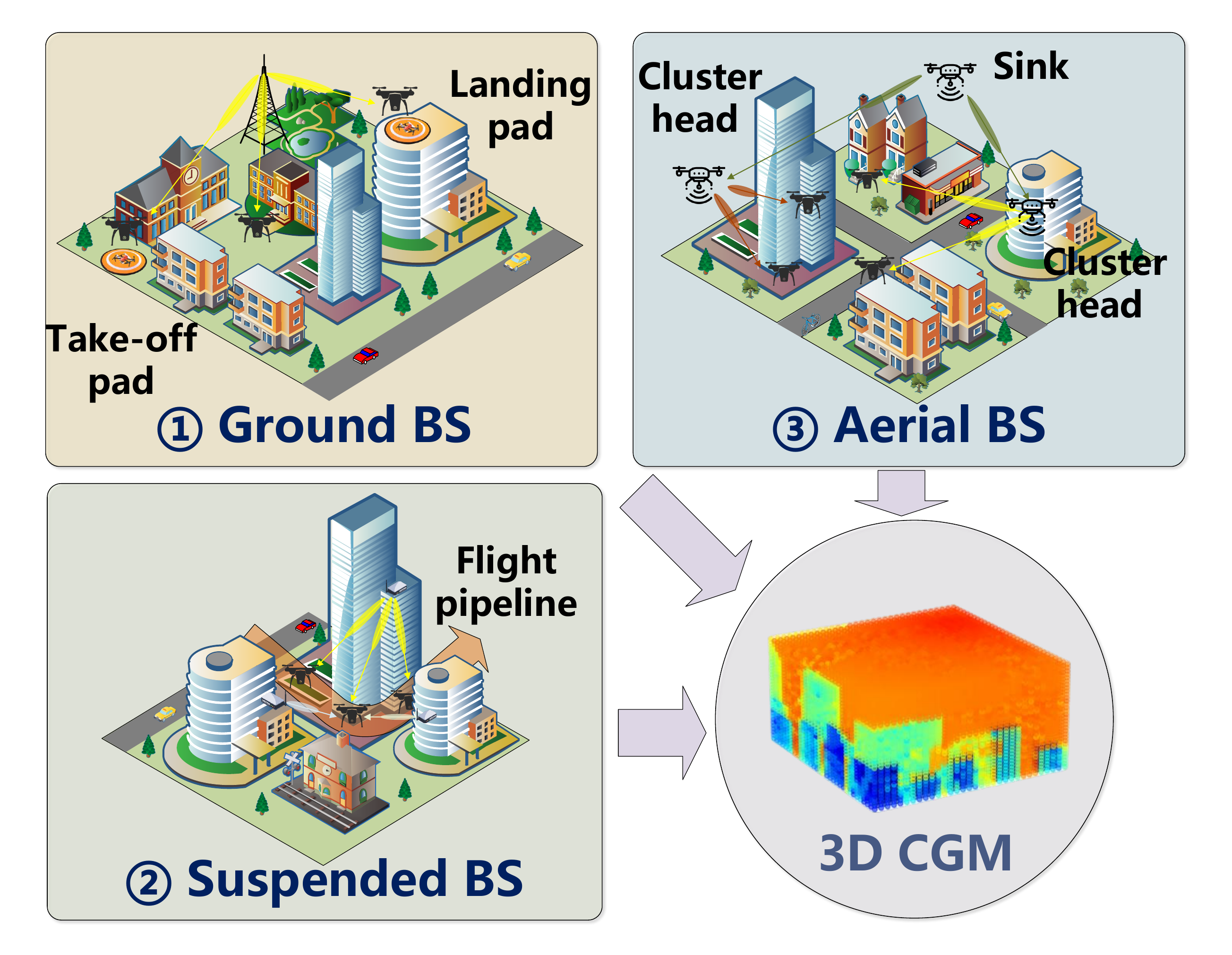}
	\caption{3D CGM inference in a urban low-altitude scenario with three types of BSs.}\label{Fig_1}
\end{figure}

As shown in Fig. \ref{Fig_1}, we consider a urban low-altitude scenario where the UAVs are controlled and served by BSs deployed with different types. The whole considered 3D region is denoted as $\mathcal{R}$, which is further partitioned into two subsets, i.e., the region $\mathcal{F}$ where the BS and UAV are permitted to be deployed and the region $\mathcal{B}$ where the buildings are occupied. It is obvious that $\mathcal{B}=\mathcal{R}\setminus \mathcal{F}$. We introduce a 3D Cartesian coordinate system, and let $\mathbf{q}=(x,y,z)\in\mathcal{R}$ and $\mathbf{o}=(x_0,y_0,z_0)$ denote the coordinates of  an arbitrary position within $\mathcal{R}$ and BS, respectively. If $\mathbf{q}\in\mathcal{F}$, it represents the coordinate of a UAV, otherwise it represents the internal coordinate of the building. Specifically, due to the existence of buildings in the urban scenarios, the BS can be primarily categorized as three types based on its deployed location:
\subsubsection{Ground BS (GBS)}
Deployed on the ground, GBS provides wide-area coverage for low-altitude UAV and ground user equipment (UE), supporting critical operations like UAV takeoff and landing.
\subsubsection{Suspended BS (SBS)}
SBS refers to the BS that is attached on the surface or deployed on the top of the buildings. By properly tuning the direction of its antenna array, SBS can provide a reliable communication in the flight pipeline between two buildings. 
\subsubsection{Aerial BS (ABS)}
ABS, implemented as low-altitude UAVs, serves as a flexible GBS or SBS, enabling dynamic deployment for applications like UAV swarm coordination or temporary coverage.
\begin{figure}[b]
	\centering
	\includegraphics[width=3in]{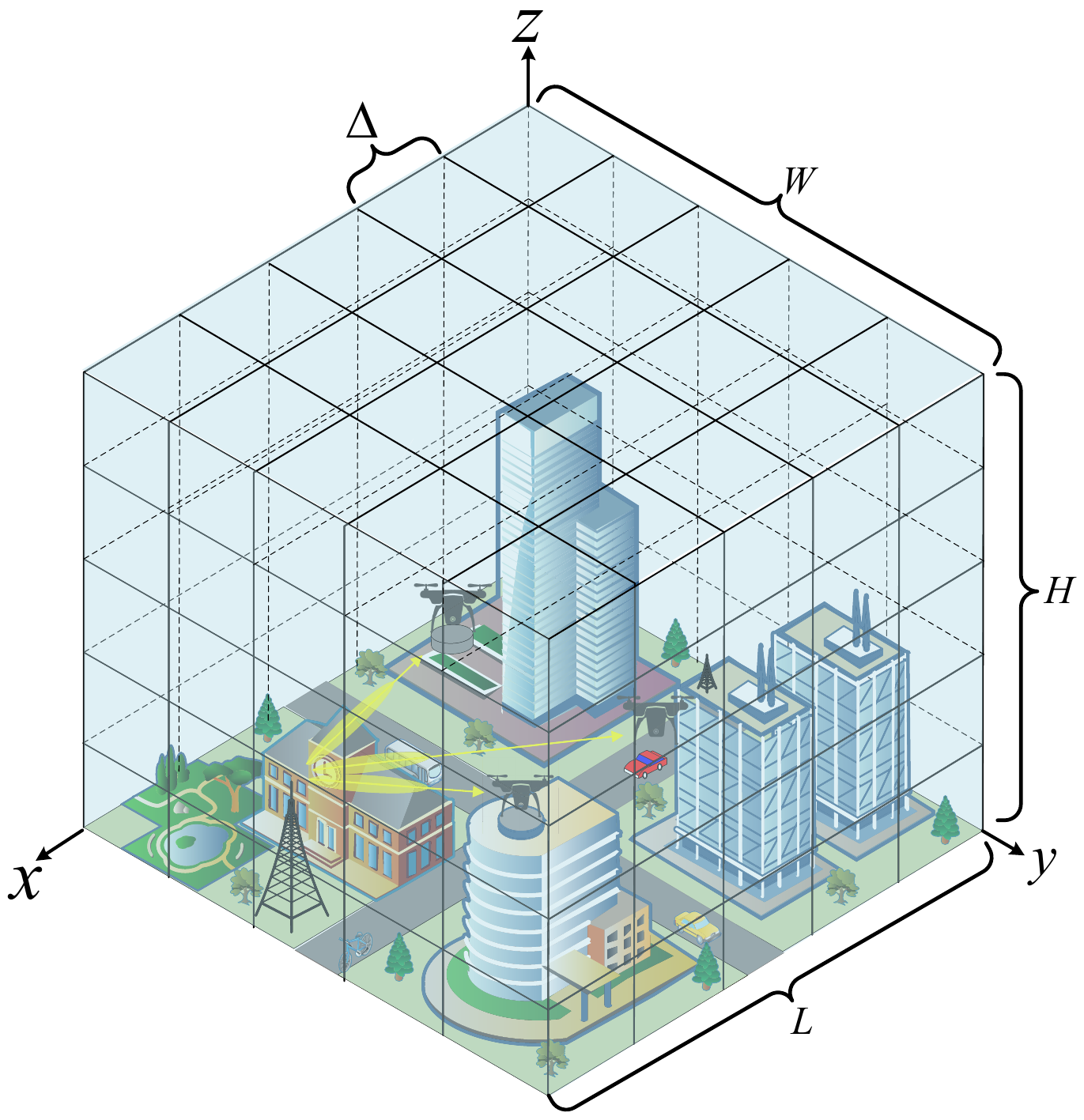}
	\caption{3D grid discretization of $\mathcal{R}$.}
	\label{fig2}
\end{figure}

In practice, the 3D urban region $\mathcal{R}$ is continuous with the length $L$, width $W$, and height $H$. However, such a continuity of the region makes it difficult to collect the wireless channel knowledge via online/offline measurements. For simplicity, we assume that $\mathcal{R}$ is shaped by a truncated 3D rectangle and then is discretized into a 3D grid model with a granularity of ${\Delta}$, as shown in Fig. \ref{fig2}. Therefore, the total number of small cubes in this grid model is given by
\begin{equation}
	\label{deqn_ex1}
		N=\lceil\frac{L} {\Delta} \rceil\times\lceil\frac{W} {\Delta} \rceil\times\lceil\frac{H} {\Delta} \rceil.
\end{equation}
Thus the number of grid cells along the length, width, and height directions are denoted as 
$N_{x}=\lceil\frac{L} {\Delta} \rceil$, $N_{y}=\lceil\frac{W} {\Delta} \rceil$, and $N_{z}=\lceil\frac{H} {\Delta} \rceil$, respectively.

Let $\mathbf{q}_{i,j,k}$ represent the coordinate of the center point of the $(i,j,k)$-th cube, denoted as $(x,y,z)_{i,j,k}$, where $i \in\{1, 2, \ldots, N_{x} \}$, $j \in\{1, 2, \ldots, N_{y} \} $, $k \in\{1, 2, \ldots, N_{z} \} $, which can be expressed as
\begin{equation}
	\mathbf{q}_{i, j, k}=( x, y, z)_{i, j, k}=( i-\frac{1} {2}, j-\frac{1} {2}, k-\frac{1} {2} ) \Delta. 
\end{equation}

\subsection{3D Channel Gain Model and Map Representation}
Generally, the channel gain $\gamma$ of a wireless link includes three components, i.e., the path loss $\gamma_{\rm PL}$, shadowing $\gamma_{\rm SH}$, and multipath fading $\gamma_{\rm MP}$\cite{ref5}, all of which are functions with respect to the locations of transmitter and receiver. Therefore, the channel gain between the BS located at $\mathbf{o}$ and $\mathbf{q}_{i,j,k}\in\mathcal{F}$ can be expressed as
\begin{equation}
\label{channel_gain}
	\gamma(\mathbf{o},{\bf q}_{i,j,k} ) = \gamma_{\mathrm{P L}} ( \mathbf{o},{\bf q}_{i,j,k} ) \gamma_{\mathrm{S H}} ( \mathbf{o},{\bf q}_{i,j,k} ) \gamma_{\mathrm{M P}} ( \mathbf{o},{\bf q}_{i,j,k} ).
\end{equation}

Let $\gamma_{\mathrm{d B}}$ denote the channel gain in dB, \eqref{channel_gain} can be re-expressed as
\begin{equation}
	\begin{aligned}
		\gamma_{\mathrm{d B}}( {\mathbf{o},\bf q}_{i,j,k} ) &= K_{\mathrm{d B}}\\
		&\quad-1 0 n_{\mathrm{P L}} \operatorname{l o g}_{1 0} ( \left\| \mathbf{o}-\mathbf{q}_{i, j, k} \right\| )\\
		&\quad+1 0 \operatorname{l o g}_{1 0} ( \gamma_{\mathrm{S H}} ( \mathbf{o},\mathbf{q}_{i,j,k} ) ) \\
		&\quad+1 0 \operatorname{l o g}_{1 0} ( \gamma_{\mathrm{M P}} ( \mathbf{o},\mathbf{q}_{i,j,k} ) )  ),
	\end{aligned}
\end{equation}
where $K_{\mathrm{d B}}$ and $n_{\rm PL}$ denote the path loss intercept and path loss exponent, respectively. Besides, if $\mathbf{q}_{i,j,k}\in\mathcal{B}$, the corresponding channel gain is set to a predefined minimum value $\gamma^{\min}_{\mathrm{dB}}$ to signify that the location is occupied by the part of a building. Therefore, $\gamma_{\mathrm{d B}}(\mathbf{o},\mathbf{q}_{i,j,k})$ within the 3D region $\mathcal{R}$ is given by
\begin{equation}
	\gamma_{\mathrm{d B}}(\mathbf{o},\mathbf{q}_{i,j,k})=
	\begin{cases}
		\gamma_{\mathrm{d B}} (\mathbf{o}, \mathbf{q}_{i,j,k}), &\mathbf{q}_{i,j,k} \in \mathcal{F}, \\
		\gamma^{\min}_{\mathrm{d B}}, &\mathbf{q}_{i,j,k} \in \mathcal{B}.
	\end{cases}
\end{equation}

The mapping relationship between $\mathbf{q}_{i,j,k}$ and $\gamma_{\mathrm{d B}}( {\mathbf{o},\bf q}_{i,j,k} )$ with the given $\mathbf{o}$ is written by
\begin{equation}
	\mathcal{M} \colon\mathbf{q}_{i,j,k} \to\gamma_{\mathrm{d B}}( {\mathbf{o},\bf q}_{i,j,k} ).
\end{equation}

Therefore, once the coordinate $\mathbf{o}$ is determined, the CGM with respect to $\gamma_{\mathrm{d B}}( {\mathbf{o},\bf q}_{i,j,k} )$ is generated, which can be expressed as
\begin{align}
	\mathbf{C}({\mathbf{o}})=&\{\left( \mathbf{q}_{1, 1, 1}, \gamma_{\mathrm{d B}}(\mathbf{o},\mathbf{q}_{1,1,1}) \right), \left( \mathbf{q}_{1, 2, 1}, \gamma_{\mathrm{d B}}(\mathbf{o},\mathbf{q}_{1,2,1})\right),\nonumber\\
	&\cdots, \left( \mathbf{q}_{N_{x}, N_{y}, N_{z}}, \gamma_{\mathrm{d B}}(\mathbf{o},\mathbf{q}_{N_{x}, N_{y},N_{z}}) \right) \}.
\end{align}

\subsection{Problem Formulation of CGM Inference}

In practice, acquiring CGMs of BSs at all locations in $\mathcal{R}$ through conventional channel measurement methods would consume a significant amount of computation and storage resources. A promising solution is to construct the CGM of the BS at a new location by using existing CGMs. These existing CGMs consist of a priori known channel gain knowledge from multiple BSs at different locations. Specifically, we assume that there are already $M$ existing CGMs, and denote the $m$-th existing CGM as $\mathbf{C}(\mathbf{o}_m),m\in\{1,\cdots,M\}$. Define $\mathcal{C}=\{\mathbf{C}(\mathbf{o}_1), \mathbf{C}(\mathbf{o}_2) \cdots, \mathbf{C}(\mathbf{o}_M) \}$ and $\mathcal{O}=\{\mathbf{o}_{1}, \mathbf{o}_{2}, \ldots, \mathbf{o}_{M} \}$ as the existing CGM set and the corresponding BS coordinate set, respectively.

We aim to learn a mapping $f$ that relates BS locations to CGMs, enabling accurate CGM inference for arbitrary $\mathbf{o} \in \mathcal{F}$ using the existing datasets $\mathcal{C}$ and $\mathcal{O}$. Once the BS coordinate is updated, we can directly obtain the corresponding CGM by such a inference model without real-time channel measurement. Specifically, for the BS location $\tilde{\mathbf{o}} \notin \mathcal{O}$, the final inferred CGM can be expressed as
\begin{equation}
\label{relation}
	\hat{\mathbf{C}}(\tilde{\mathbf{o}})=f \bigl( \tilde{\mathbf{o}} \bigr).
\end{equation}

\section{3D CGM Inference Based on 3D-CGAN}
In this section, we propose a method based on 3D-CGAN to establish the above mentioned CGM inference procedure. Specifically, existing CGMs $\mathcal{C}$ and BS location coordinates $\mathcal{O}$ are utilized as the training dataset to train the model. Once the training is accomplished, the model is able to output the inferred CGM $\hat{\mathbf{C}}(\tilde{\mathbf{o}})$ corresponding to updated coordinate of BS $\tilde{\mathbf{o}}$.

\subsection{Dataset Construction}
The dataset utilized for model training is composed of existing CGMs in $\mathcal{C}$, which are generated by actual environment. Specifically, we employ data simulated through the ray tracing method as a proxy for the real data, which are generated using the software Remcom Wireless InSite. The dataset is constructed with the following steps: \romannumeral 1) Establish a model of $\mathcal{R}$ in software Wireless InSite based on the standardized urban model ITU-R Rec.P.1410 proposed by the International Telecommunication Union (ITU-R)\cite{ref11}; \romannumeral 2) Configure the antennas, waveform, and other relevant parameters; \romannumeral 3) Based on the previously established 3D grid model, the receivers are deployed at the central coordinates of each cubic cell; \romannumeral 4) Place transmitters at the locations where data collection is planned; \romannumeral 5) Select the types of results of interest and conduct the simulation. 

When initializing the considered 3D urban low-altitude region, some parameters in terms of buildings should be defined in advance. Specifically, the ratio of building footprint to the total area, denoted by $\alpha$, is set to 0.5, the average number of buildings per square kilometer $\beta$ is set to 300, and the standard deviation  of the building height distribution $\gamma$ is set to 50 meters. The height of the building $h$ can be obtained through its probability density function $P(h)$, i.e.,
\begin{equation}
\label{height}
P ( h )=\frac{h} {\gamma^{2}} e^{( \frac{-h^{2}} {2 \gamma^{2}} )}.
\end{equation}

The size of $\mathcal{R}$ is set to $256m\times256m\times128m$, which is discretized into a grid consisting of $32\times32\times32$ cuboids. The predefined minimum value $\gamma^{\min}_{\mathrm{dB}}$ is set to -250 dB in the software simulation. Based on the initialized parameters $\alpha$, $\beta$, $\gamma$, and $h$, we place 20 buildings within the region and select a set of heights for them with \eqref{height}. Place receivers at the center points of each cuboid, and select appropriate locations within $\mathcal{F}$ to place transmitters. Finally, we collect 950 BS coordinates along with their corresponding CGMs.

\begin{figure}[t]
	\centering
	\includegraphics[width=3.5in]{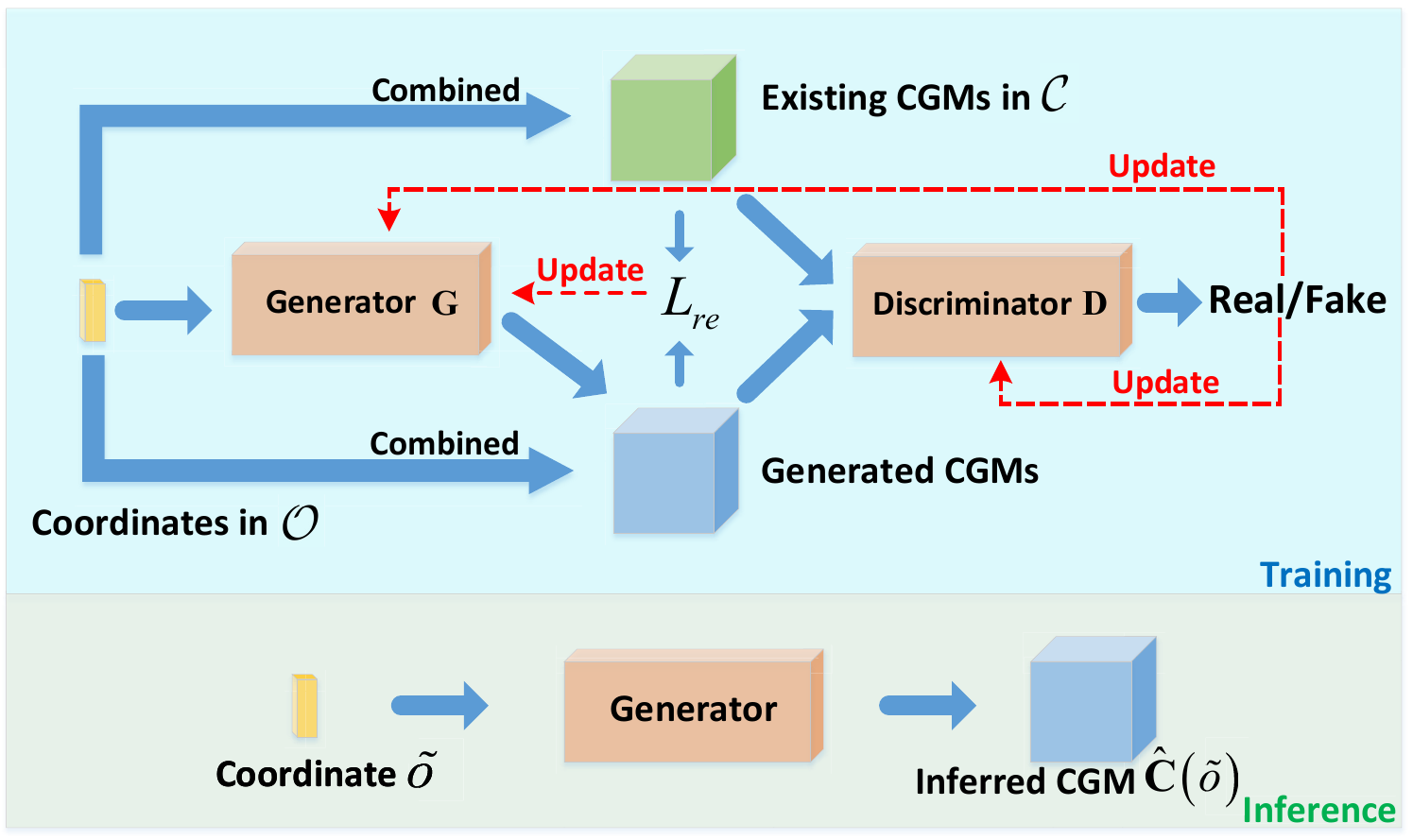}
	\caption{The CGM training and inference phases of the proposed 3D-CGAN.}
	\label{process}
\end{figure}
\begin{figure*}[b]
	\centering
	\includegraphics[width=\textwidth]{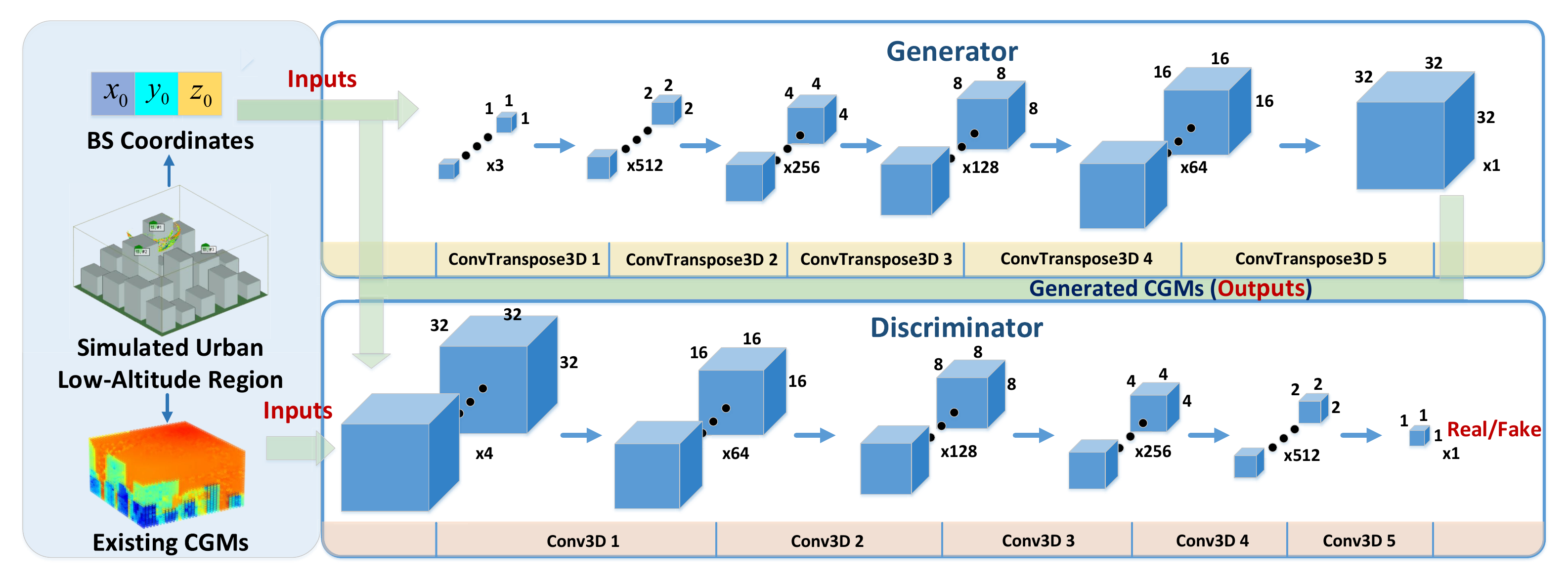}
	\caption{The block diagram of the 3D-CGAN for CGM inference.}
	\label{gan}
\end{figure*}

\subsection{Design of The 3D-CGAN}
The proposed 3D-CGAN consists of a generator $\mathbf{G}$ and a discriminator $\mathbf{D}$\cite{ref12}, which carries out the 3D CGM inference by two phases: the training phase and the inference phase, as illustrated in Fig. \ref{process}. Specifically, the coordinates of the BSs in $\mathcal{O}$ are uniquely paired with their corresponding existing CGMs in $\mathcal{C}$. During the training phase, the generator takes the coordinates as three-channel inputs and produces what we call the generated CGMs as outputs. The discriminator has two types of inputs: the first involves four-channel data containing existing CGMs with coordinates, and the second involves combining generated CGMs with coordinates. Based on these inputs, the discriminator has two outputs, i.e., $\mathbf{D}(\mathbf{C}(\mathbf{o})|\mathbf{o})\in(0,1)$ and $\mathbf{D}(\mathbf{G}(\mathbf{o})|\mathbf{o})\in(0,1)$, respectively, which determine whether the CGM combined with the coordinate is the existing CGM or the generated one. Subsequently, the generator and discriminator are updated based on such two values. Additionally, the difference between the generated and existing CGMs is used to update $\mathbf{G}$. Specifically, the loss of $D$ is denoted as $\mathbb{E}_{\mathbf{C}({\mathbf{o}}),\mathbf{o}} \left[ ( \mathbf{D}(\mathbf{C}({\mathbf{o}}) | \mathbf{o})-1 )^{2} \right]+\mathbb{E}_{\mathbf{o}} \left[ ( \mathbf{D} ( \mathbf{G}( \mathbf{o} ) | \mathbf{o} ) )^{2} \right]$, and the loss of $\mathbf{G}$ is denoted as $\mathbb{E}_{\mathbf{o}} \left[ ( \mathbf{D} ( \mathbf{G}( \mathbf{o}) | \mathbf{o} )-1 )^{2} \right]+\lambda_{re} L_{re}$, where $L_{re}=\mathbb{E}_{\mathbf{C}({\mathbf{o}}), \mathbf{o}} \left[\frac{1}{N} \| \mathbf{G}( \mathbf{o} )-\mathbf{C}({\mathbf{o}}) \|_{2}^{2} \right]$ denotes the reconstruction loss, $\lambda_{re}$ denotes the weight of $L_{re}$. After training phase, the model can learn the mapping $f$ in \eqref{relation} between the coordinates and their corresponding CGMs, as well as the environmental information within $\mathcal{R}$, such as the locations of buildings. During the inference phase, by providing a new coordinate $\tilde{\mathbf{o}}$ to the generator, we can obtain an inference $\hat{\mathbf{C}}(\tilde{\mathbf{o}})$. The details of the proposed 3D-CGAN are illustrated in Fig. \ref{gan}, which can be summarized as follows:
\begin{itemize}
	\item {\bf Generator}: The coordinates $\{\mathbf{o}_{m}\}$ serve as inputs to the generator. The generator consists of five layers of 3D transposed convolutional layers, which progressively upsample a low-dimensional input into a $32\times32\times32$ CGM. Each layer uses a $4\times4\times4$ kernel, stride 2, and padding 1. Except for the last layer, each of the first four transposed convolutional layers is followed by a 3D batch normalization layer and a ReLU activation function. The data transforms from 3 channels to 512 channels and then gradually reduces to 1 channel, while the $1\times1\times1$ data progressively expands to a $32\times32\times32$ output.

	\item {\bf Discriminator}: The discriminator consists of five layers of 3D convolutional layers, which progressively downsample the input to a scalar output. The convolution kernel size is also designed as $4\times4\times4$, with a stride of 2 and padding of 1. Except for the last layer, each of the first four convolutional layers is followed by a 3D batch normalization layer and a LeakyReLU activation function. After the last convolutional layer, a Sigmoid activation function is applied to indicate whether the input data is real. The input, consisting of 4-channel data that includes 3D data and coordinates, is gradually transformed into 512 channels and then output as one-channel data. The shape of the data is also changed from $32\times32\times32$ to $1\times1\times1$.
\end{itemize}

\begin{figure*}[t] 
	\centering
	% 第一排图片
	\begin{subfigure}[b]{0.22\textwidth}
		\includegraphics[trim=80 70 90 100, clip, width=\textwidth]{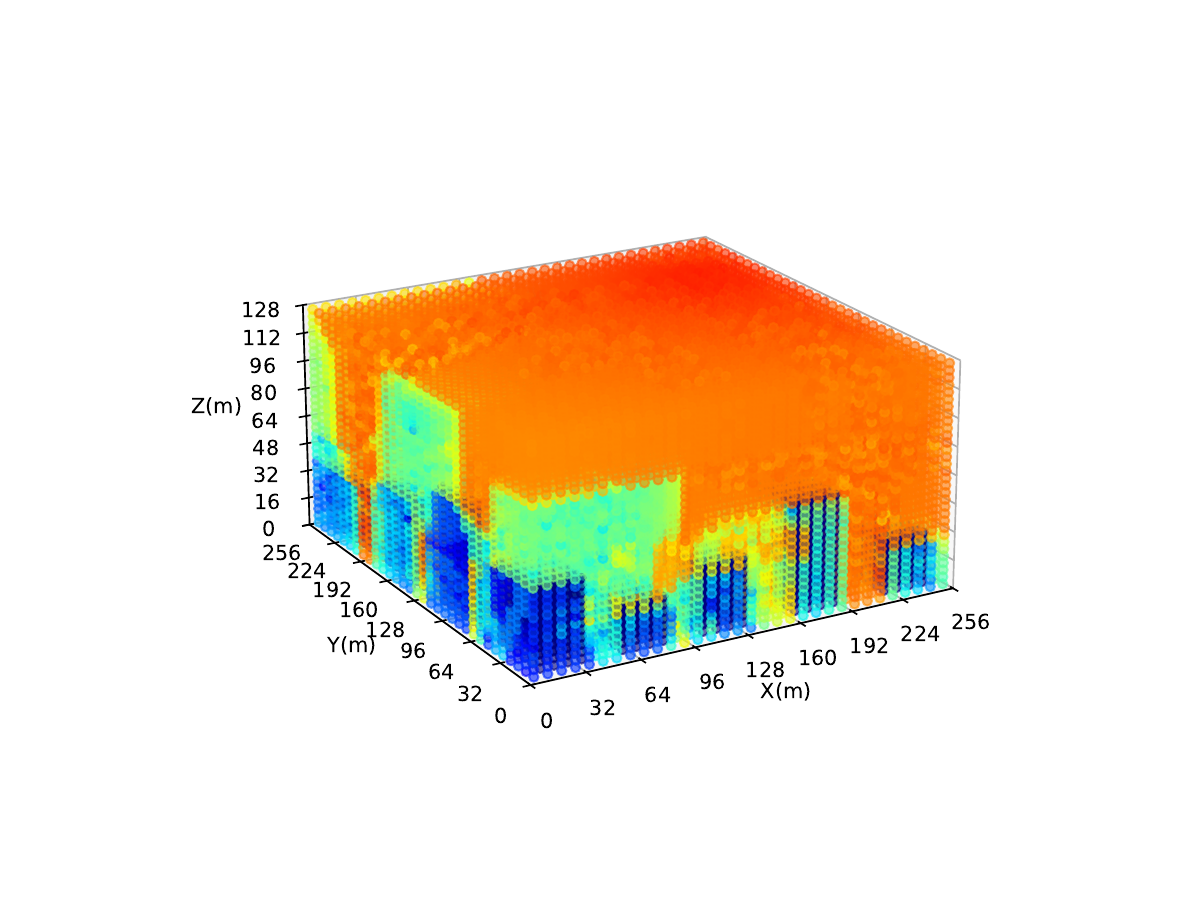}
		\caption{Ground-truth CGM}
		\label{1a}
	\end{subfigure}
	\hfill % 添加一些间距
	\begin{subfigure}[b]{0.22\textwidth}
		\includegraphics[trim=80 67 90 100, clip, width=\textwidth]{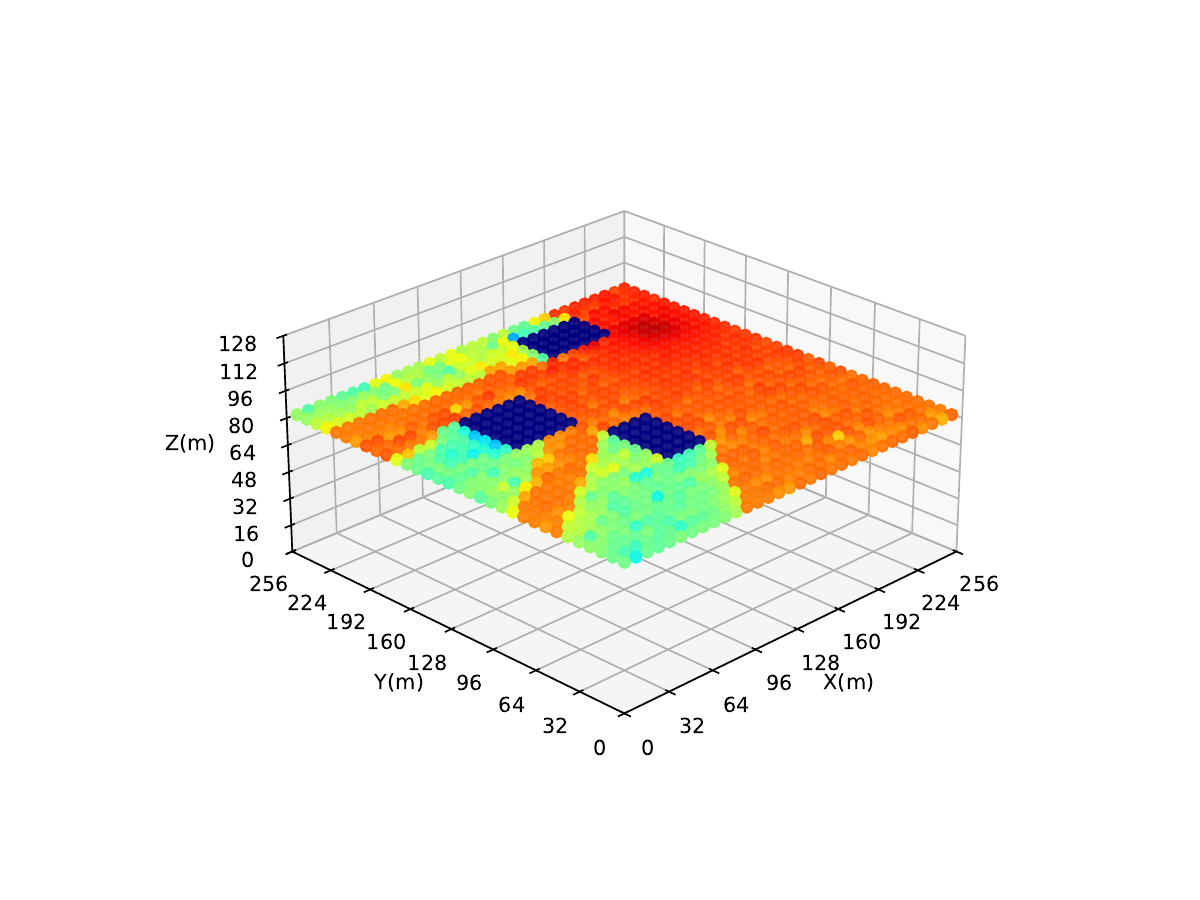}
		\caption{$Z=82m$}
	\end{subfigure}
	\hfill % 添加一些间距
	\begin{subfigure}[b]{0.22\textwidth}
		\includegraphics[trim=80 70 90 100, clip, width=\textwidth]{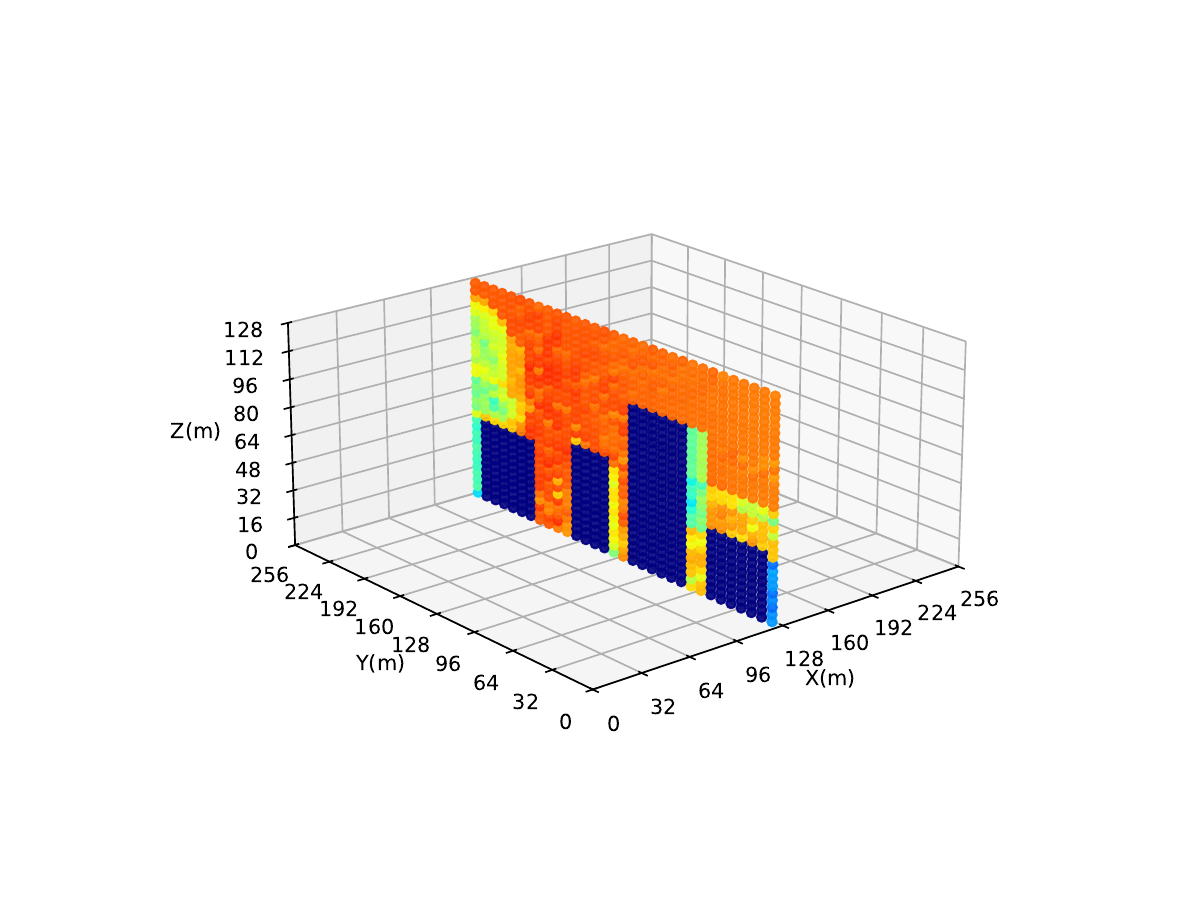}
		\caption{$X=124m$}
	\end{subfigure}
	\hfill % 添加一些间距
	\begin{subfigure}[b]{0.26\textwidth}
		\includegraphics[trim=45 70 67 100, clip, width=\textwidth]{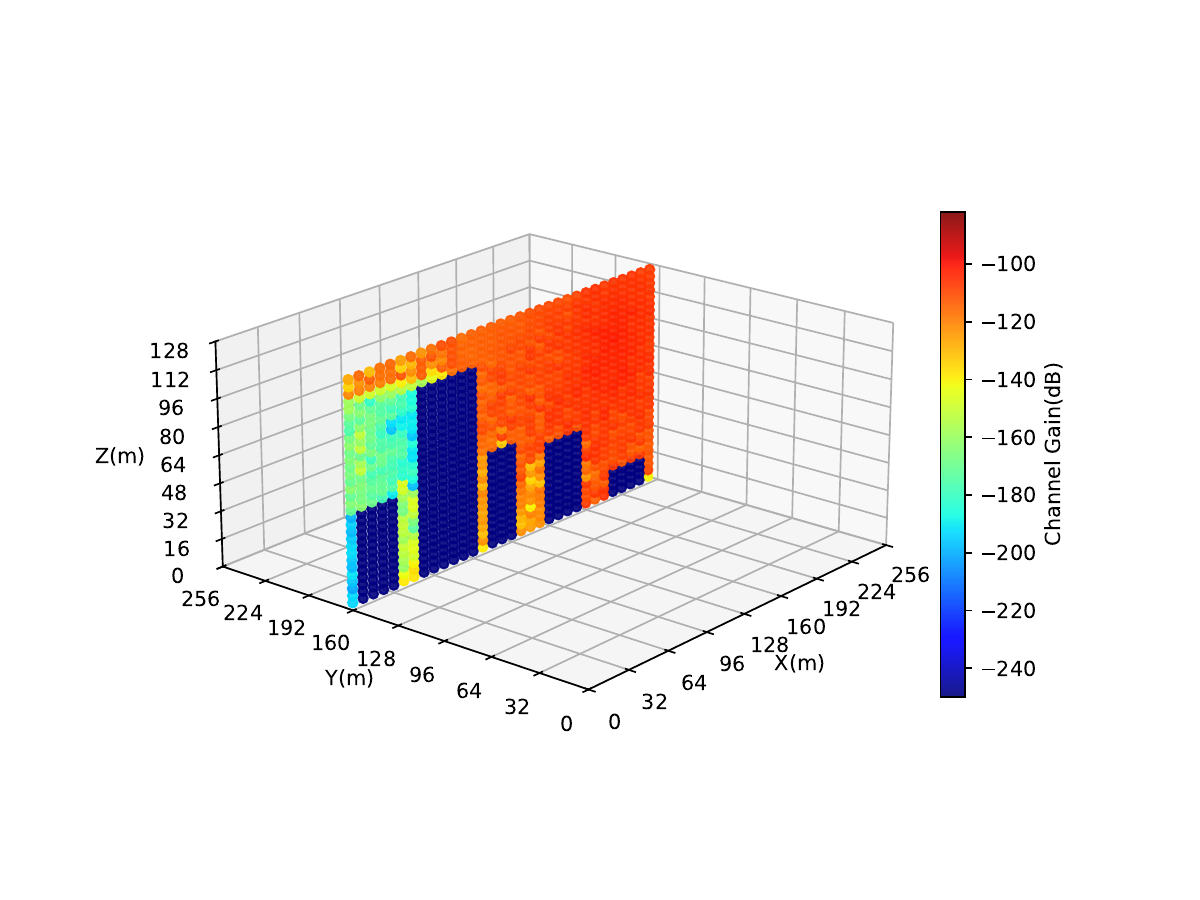}
		\caption{$Y=164m$}
		\label{1d}
	\end{subfigure}
	% 第二排图片
	\begin{subfigure}[b]{0.22\textwidth}
		\includegraphics[trim=80 70 90 100, clip, width=\textwidth]{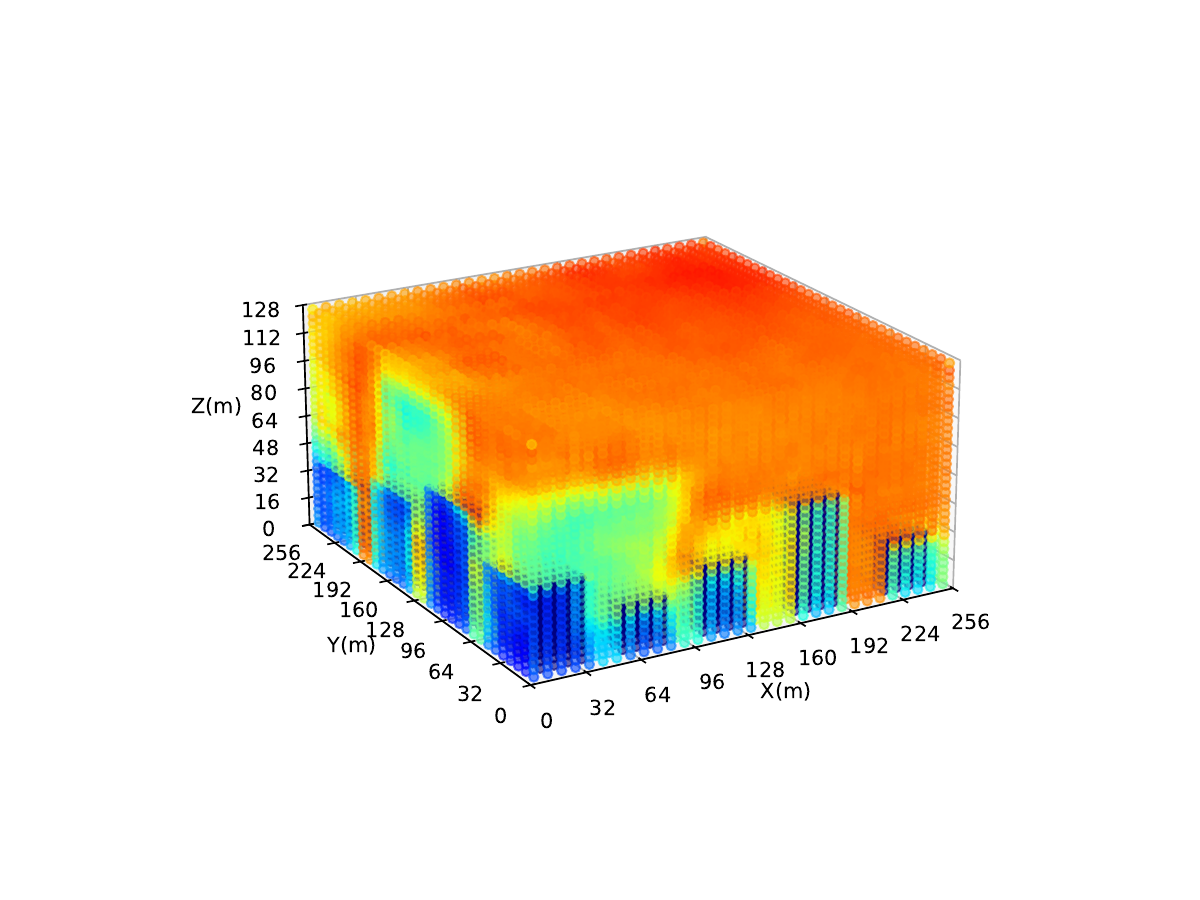}
		\caption{Inferred CGM}
	\end{subfigure}
	\hfill % 添加一些间距
	\begin{subfigure}[b]{0.22\textwidth}
		\includegraphics[trim=80 67 90 100, clip, width=\textwidth]{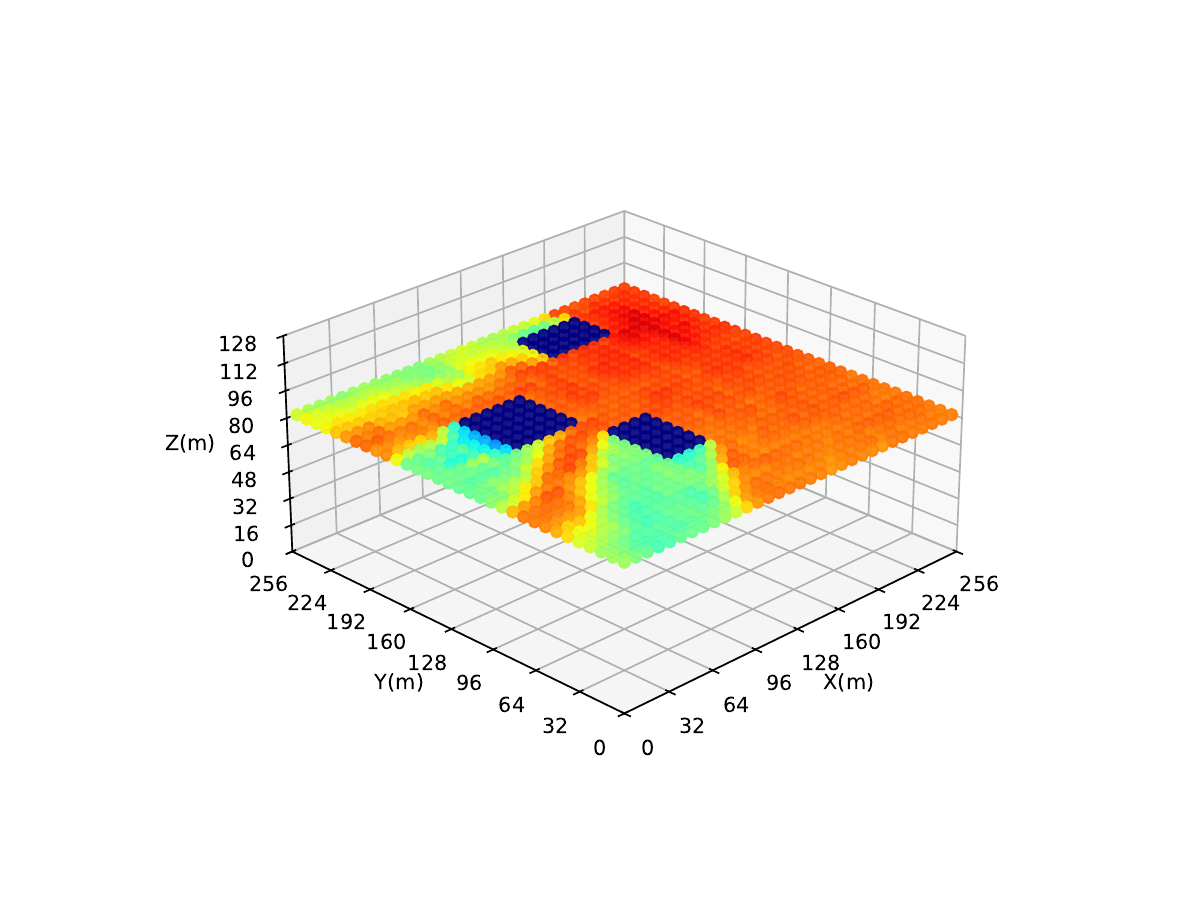}
		\caption{$Z=82m$}
	\end{subfigure}
	\hfill % 添加一些间距
	\begin{subfigure}[b]{0.22\textwidth}
		\includegraphics[trim=80 70 90 100, clip, width=\textwidth]{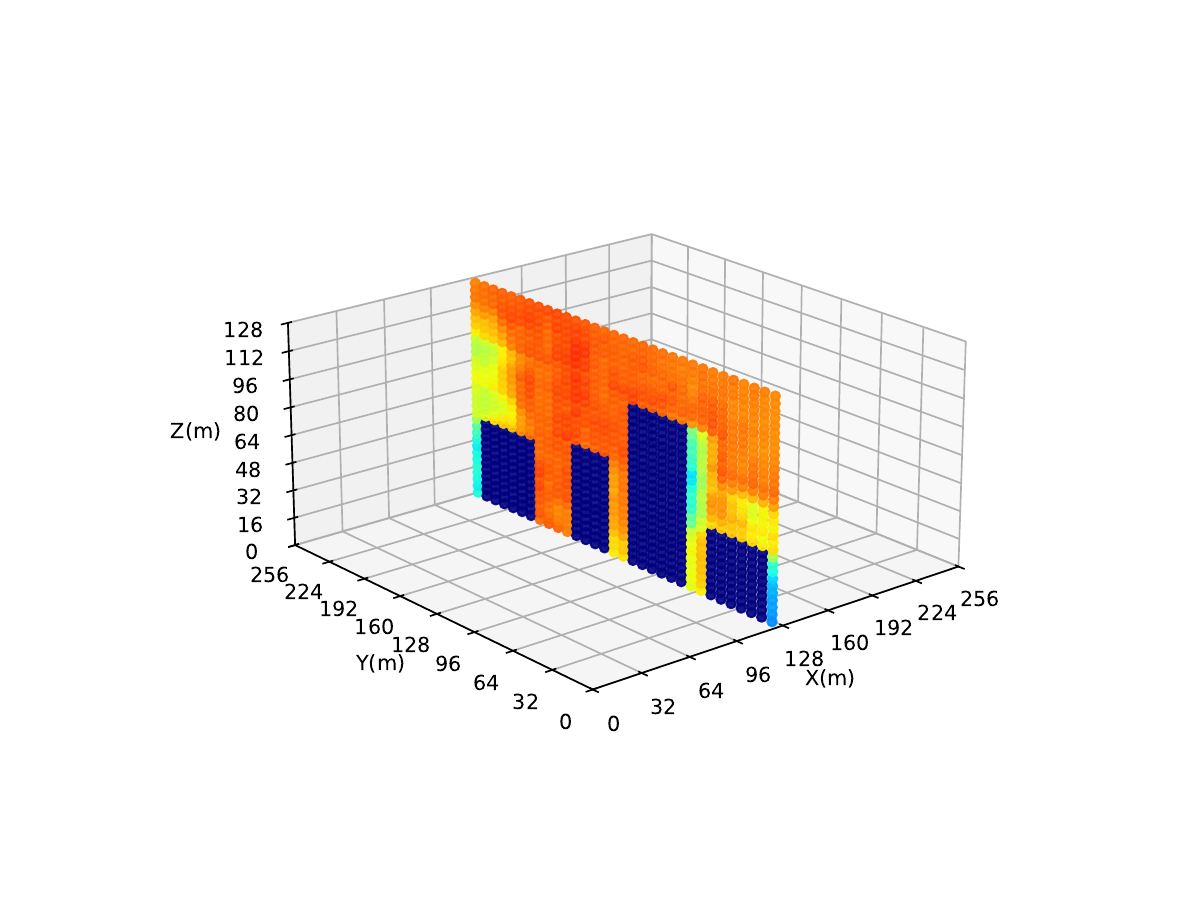}
		\caption{$X=124m$}
	\end{subfigure}
	\hfill % 添加一些间距
	\begin{subfigure}[b]{0.26\textwidth}
		\includegraphics[trim=45 70 67 100, clip, width=\textwidth]{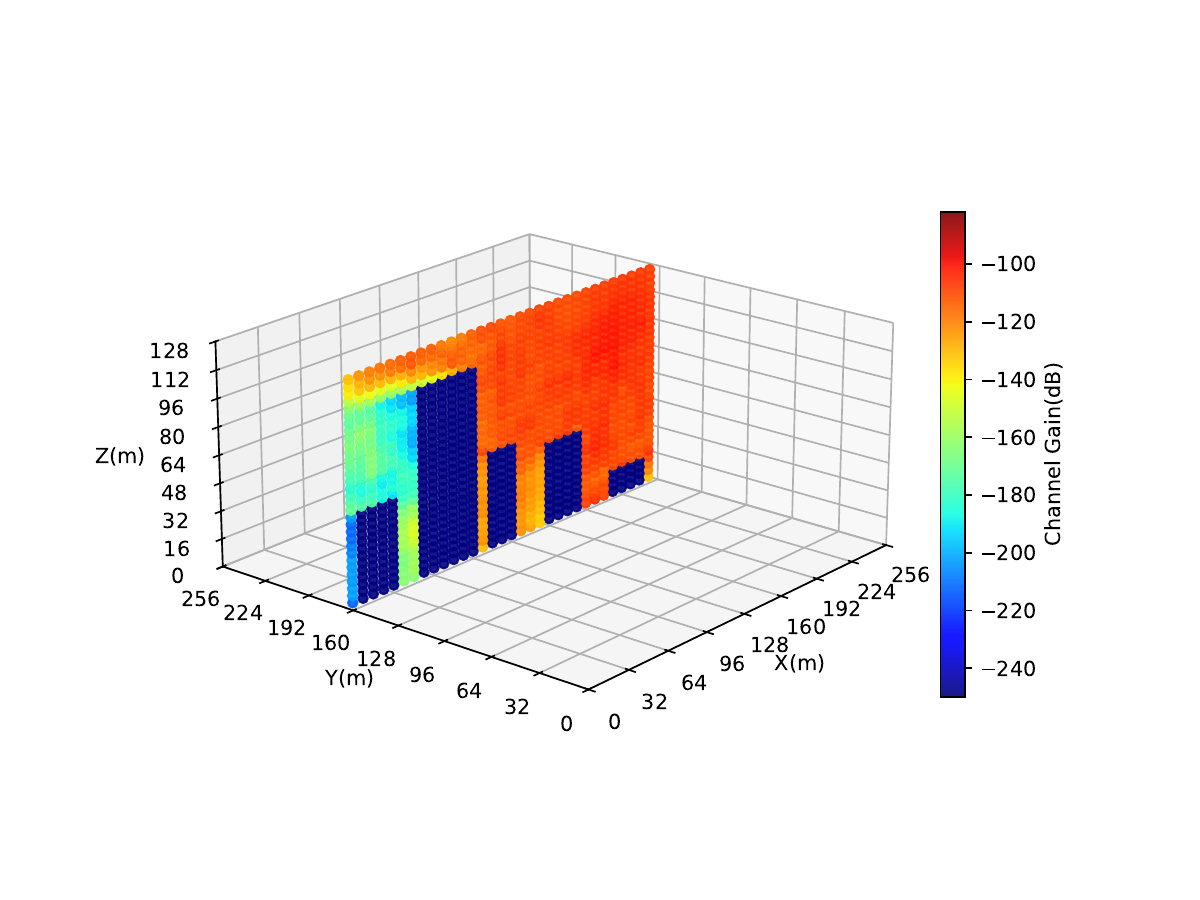}
		\caption{$Y=164m$}
	\end{subfigure}
	\caption{3D visualization comparison between ground-truth CGM and inferred CGM.}
		\label{CGM}
\end{figure*}

\subsection{Evaluation Metric}
After training the network, the trained model can be directly utilized for rapid inference of the CGM for BS at an arbitrary unknown location. The gap between the inferred CGM and the ground-truth CGM should be measured to determine the inference accuracy of the proposed 3D-CGAN. To assess the accuracy of the inferred CGMs, we use the average mean square error (AMSE), which measures the squared difference between inferred and ground-truth channel gains across all grid points as
\begin{multline}
	{AMSE} = \frac{1}{I \times N} \sum_{m=1}^{I} \sum_{\mathbf{q}_{i,j,k} \in \mathcal{R}} \\\left( \gamma_{\mathrm{d B}}(\mathbf{o}_{m},\mathbf{q}_{i,j,k}) - \hat{\gamma}_{\mathrm{d B}}(\mathbf{o}_{m},\mathbf{q}_{i,j,k}) \right)^2,\label{AMSE}
\end{multline}
where $I$ represents the number of CGMs that are about to be inferred, and $\hat{\gamma}_{\mathrm{d B}}(\mathbf{o}_{m},\mathbf{q}_{i,j,k})$ denotes the inferred channel gain.

\section{Performance Evaluation}
In this section, we utilize the constructed dataset to train the proposed 3D-CGAN network and employ the trained model to infer the CGMs on the test set. We compare the inferred CGM results with those obtained by a benchmark scheme to validate the effectiveness of the proposed method.

\subsection{Training Settings}
To conduct network training and effectively evaluate the inference performance of the proposed scheme, the 950 sets of data are divided into two parts: 900 sets for training and 50 sets for testing. Each set of data contains a BS coordinate and its corresponding CGM. Training is conducted multiple times based on different quantities of training data. To ensure the validity of the evaluation, the training data are randomly selected from the 900 sets, while keeping the 50 sets of test data unchanged. The model training is based on the Adam optimizer, with the initial learning rate set to 0.0002. The batch size varies depending on the amount of training data. Based on the loss values during training, models with lower losses are saved.

\subsection{Performance Analysis}

Once the model training is accomplished, only the generator is used during the inference stage. By inputting the 50 coordinates from the test set into generator, 50 corresponding inferred CGMs can be obtained. Fig. \ref{CGM} shows the comparison between the ground-truth CGM and the inferred CGM of the BS located at $\tilde{\mathbf{o}}=(220, 200, 90)$. The color bar represents the different values of channel gain, which approximately varies from -250 $\mathrm{d B}$ to -70 $\mathrm{d B}$. Specifically, the first row (Fig. \ref{CGM}(a) to Fig. \ref{CGM}(d)) shows the ground-truth CGM. Besides, Fig. \ref{CGM}(b), \ref{CGM}(c), \ref{CGM}(d) are the cross-sectional views of Fig. \ref{CGM}(a) when sliced by the planes $Z=82m$, $X=124m$, and $Y=164m$. The second row (Fig. \ref{CGM}(e) to Fig. \ref{CGM}(h)) depicts the corresponding inferred CGM. It can be seen that the  blue cuboids in the figure represent the buildings that have minimum channel gain value, while the others indicate locations with higher channel gain.  As expected, the generator is able to accurately capture location information and generate the corresponding CGM. Since the model can distinguish the link blockage caused by the buildings. 

\begin{figure}[t]
	\centering
	\includegraphics[trim=35 20 50 51, clip, width=3.2in]{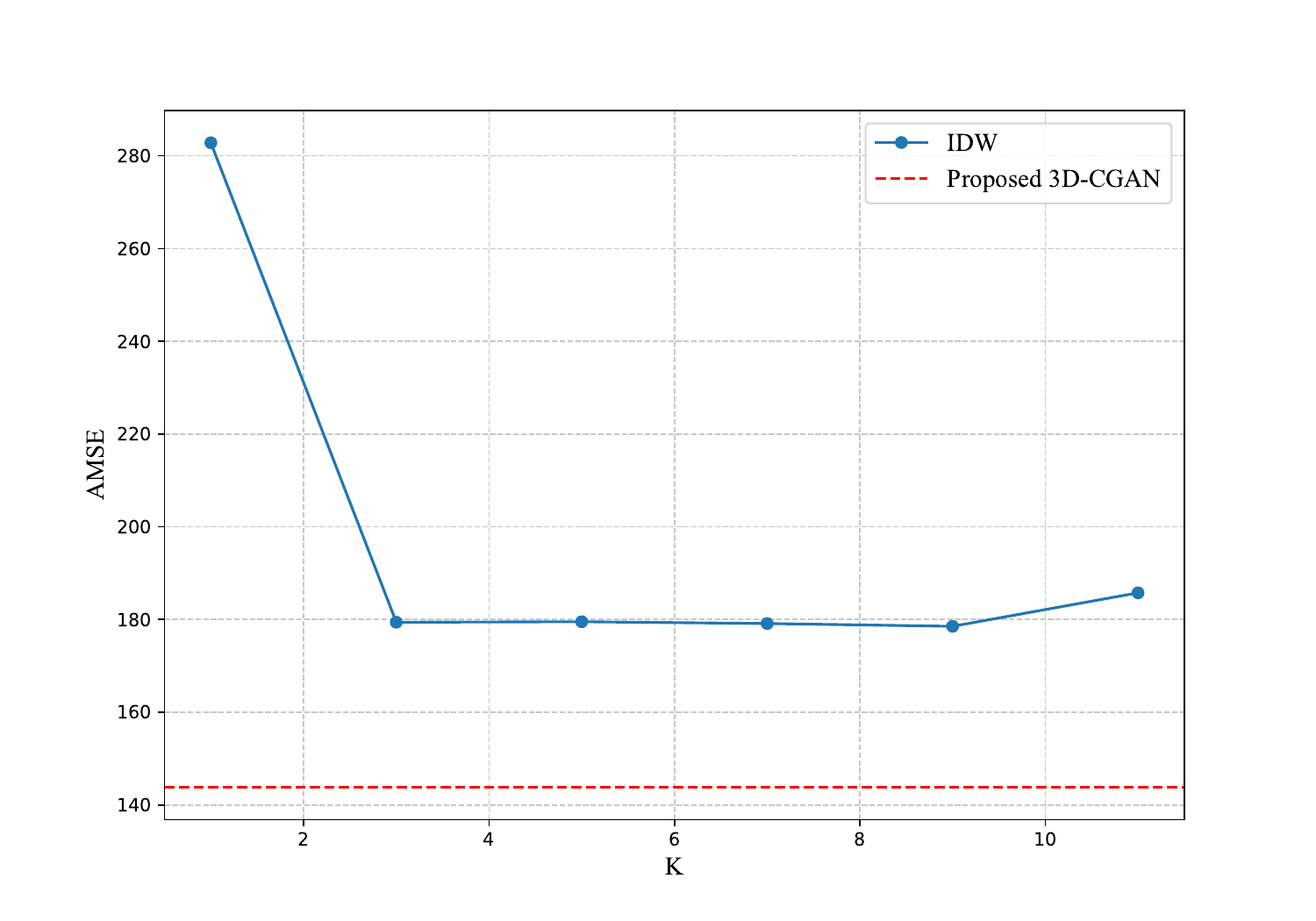}
	\caption{The AMSE comparison of the proposed 3D-CGAN and IDW on CKM inference with different $K$. }
	\label{varyk}
\end{figure}

To measure the performance of the proposed 3D-CGAN method, we use \eqref{AMSE} to calculate AMSE of the generator on the test set. Simultaneously, we employ the inverse-distance-weighted (IDW)-based CGM inference method as a benchmark, by which the CGM is obtained as follows
\begin{equation}
	\hat{\mathbf{C}}(\tilde{\mathbf{o}})=\frac{\sum_{m=1}^{K} \frac{\mathbf{C}(\mathbf{o}_m)} {d_{m}^{p}}} {\sum_{t=1}^{K} \frac{1} {d_{t}^{p}}},
\end{equation}
where $d$ represents the distance from the location corresponding to the existing CGM to the target location, $K$ is the number of BSs that are nearest to the BS located at $\tilde{\mathbf{o}}$ in $\mathcal{O}$, and $p$ is the power exponent. The idea of IDW is to infer the target CGM by calculating the weighted sum of other CGMs. Fig. \ref{varyk} displays the AMSE of the proposed method and the IDW CKM inference with different $K$. The proposed 3D-CGAN has an AMSE value of 143.79 $\mathrm{d B}^2$, which is a 34.75 $\mathrm{d B}^2$ reduction compared to the IDW (178.54 $\mathrm{d B}^2$ with $K=9$). In contrast to the IDW, the trained model in 3D-CGAN has the option to not retain the existing CGMs, thereby reducing storage size to 20\% of that of the IDW.

\begin{figure}[t]
	\centering
	\includegraphics[trim=0 10 0 0, clip, width=3.2in]{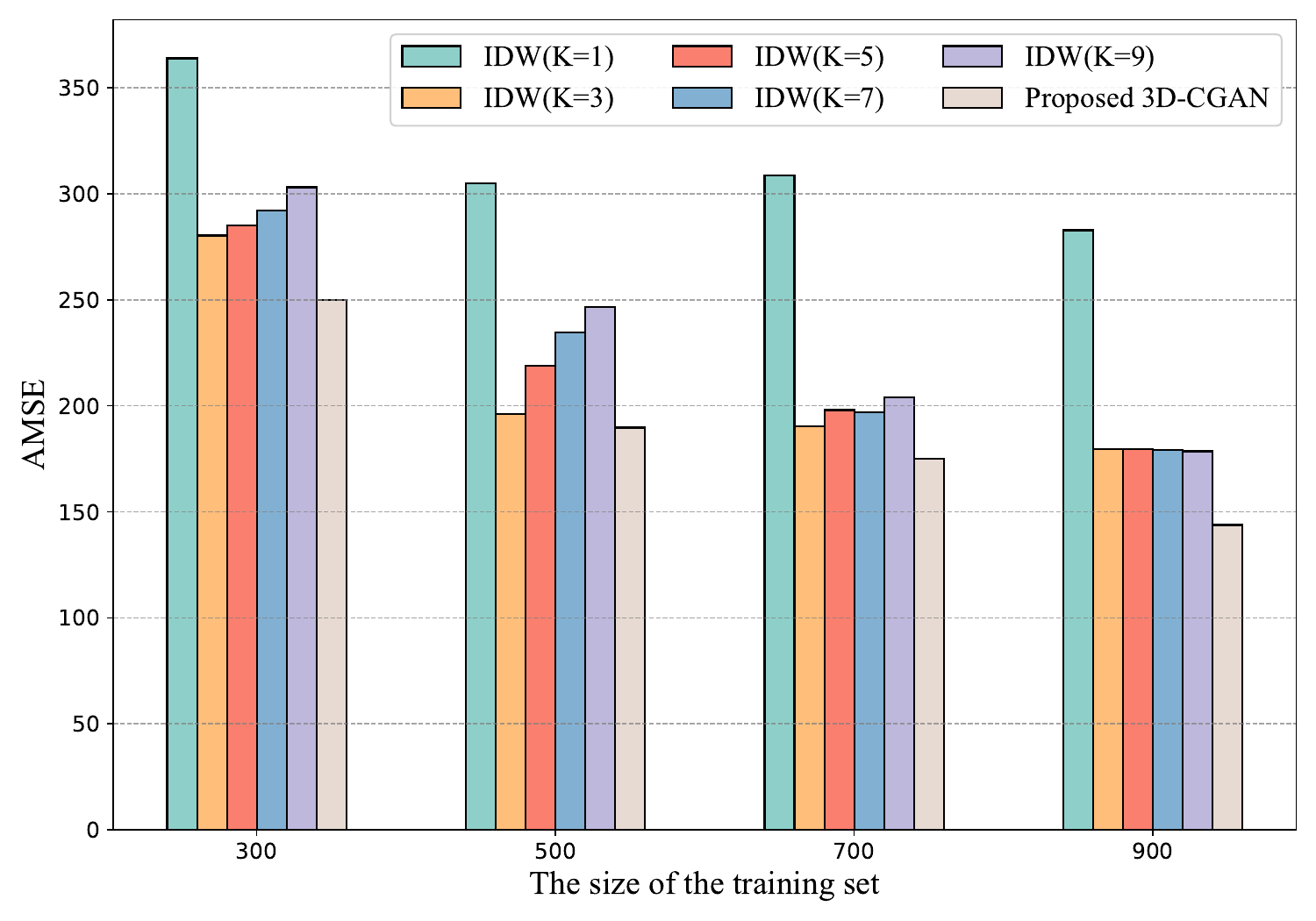}
	\caption{The AMSE comparison of the proposed 3D-CGAN and IDW with different training set size.}
	\label{bar}
	\vspace{-5mm}
\end{figure}

Fig. \ref{bar} further compares the AMSE of the proposed 3D-CGAN and IDW by fixing the test set while changing the size of the training set.  The model is trained several times. It is evident that as the size of the training set increases, the AMSE of both methods decreases. This is because a larger training set size improves the diversity of BS locations, helping the model better learn the mapping $f$. Besides, with the same size of the training set, the proposed 3D-CGAN method consistently outperforms the IDW, which indicates that the proposed method can achieve the same CGM inference performance as the IDW with fewer training data acquisition.

\section{Conclusion}
This paper proposed a 3D CGM inference method for arbitrary unknown BS locations in a urban low-altitude scenario based on 3D-CGAN. Based on the existing CGMs, this method explored the implicit relationships among the BS location, the corresponding CGM, the existing CGMs, and the physical environment. Therefore, by utilizing the trained model, it is possible to directly infer the CGM of a BS solely from its coordinate. The simulation results compared the inference errors of the proposed 3D-CGAN and IDW with different numbers of training samples, which demonstrated the effectiveness of the proposed method in inferring CGM guided by BS coordinate without additional measurement.

\vfill

\end{document}